\documentclass[11pt]{article}

\begin{document}
\newcommand {\be}{\begin{equation}}
\newcommand {\ee}{\end{equation}}
\newcommand {\bea}{\begin{array}}
\newcommand {\cl}{\centerline}
\newcommand {\eea}{\end{array}}
\renewcommand {\theequation}{\thesection.\arabic{equation}}
\renewcommand {\thefootnote}{\fnsymbol{footnote}}
\newcommand {\newsection}{\setcounter{equation}{0}\section}
\csname @addtoreset\endcsname{equation}{section}

\def\simlt{\stackrel{<}{{}_\sim}}
\def\simgt{\stackrel{>}{{}_\sim}}
\def\IP{\relax{\rm I\kern-.18em P}}
\def\sugra{supergravity }
\def\com{commutative }
\def\nc{noncommutative }
\def\ncy{noncommutativity }
\def\NCOS{noncommutative open string}
\def\ba{\begin{eqnarray}}
\def\ea{\end{eqnarray} }
\def\o{\over}  
\def \a {\alpha }
\def\th{\theta }
\def\s{\sigma}
\def\t{\tau }
\def\p{\partial }
\baselineskip 0.6 cm
\begin{flushright}
hep-th/0101045
\vskip 5mm
\end{flushright}
\begin{center}
{\LARGE{Noncommutative  Open String Theories and Their Dualities }}
\vskip 2mm

{\bf {\large{M. M. Sheikh-Jabbari}}}  

{\it The Abdus Salam International Center for Theoretical Physics\\
 Strada Costiera, 11. 34014, Trieste, Italy}\\
{\tt  jabbari@ictp.trieste.it }\\

\end{center}

\vskip 2mm

The recently found non-critical open string theories is reviewed. These
open strings, \NCOS{} theories (NCOS), arise as consistent quantum
theories describing the low energy theory of D-branes in a background
electric B-field in the critical limit. Focusing on the D3-brane case, we
construct the most general (3+1) NCOS, which is described by four
parameters. We study S and T -dualities of these theories and argue the
existence of a U-duality group{\footnote{This is the talk presented
to the conference {\it Brane New World and Noncommutative Geometry},
Torino, Villa Gualino,(Italy) October, 2000. Based on the hep-th/0006202 
and 0009141.}.


\newsection{Introduction and Review }

It has been recently shown that \nc spaces, with the coordinates
satisfying
\be
[x^{\mu},x^{\nu}]=i\theta^{\mu\nu} \ ,
\ee
arise as the worldvolume theory of $Dp$-branes in a $B_{\mu\nu}$-field
background (for a review see \cite{SW}).
Also it has been shown that the low energy theory of such branes is a
NC$U(1)$ SYM theory. Similar to the \com case, 
when $N$ number of such branes coincide this gauge symmetry is enhanced
NC$U(n)$. For the $\theta^{0\mu}=0$, the magnetic B-field, cases the large
$n$ limit of \nc gauge theories can be studied through a generalized form
of gravity/gauge theory correspondence \cite{AOS}.

The case $\theta^{0\mu}\neq 0$, the electric B-field, seems to be more
interesting because, $i)$ as a field theory they are not unitary and hence
not a well-defined quantum {\it field} theory \cite{{Gom},{SST}}. 
$ii)$ for the 3+1 dimensional case, they appear as the strong coupling
limit of NC$U(n)$ theory \cite{{GMMS},{Ganor}}. The latter can be
understood noting the Montonen-Olive duality and also the fact that 
the B-field background can be thought as the non-zero electric and
magnetic background fields of the $U(n)$ gauge theory. Under the S-duality
the electric and magnetic fields are replaced while the coupling goes
to one over itself, i.e. in the strong coupling a magnetic \nc theory is
mapped into an electric one, though in the critical electric field
limit \cite{GMMS}. Intuitively, the open strings attached to
D-branes+B-field are electric dipoles \cite{Dip}, and the critical $E$
limit occurs when the tension of the open strings becomes equal to the
force on this electric dipole \cite{SST}.

One can show that in this limit the closed strings of the bulk gravity
theory become infinitely massive (compared to open strings) and hence they
decouple from the dynamics of the theory. So in the end, in the critical
$E$ limit, we remain with a theory of \NCOS{ }, NCOS, without any closed
string states \cite{{SST},{GMMS}}. The above argument is quite general
and is true
for $Dp$-branes, $p=1,2,..,5$, however here we only focus on the $p=3$
case. It is interesting to study the dualities, S and T, for this case.

In the next section we construct the most general 3+1 \NCOS s and then
study the behaviour of these theories under the usual IIB S-duality
transformation. We show that under S-duality in general a NCOS is mapped
into another NCOS . However, there are some special cases for which a NCOS
goes to a NCSYM \cite{RS1}.

In section 4, we compactify the 3+1 NCOS on a $T^2_{\theta}$, so that the
magnetic component of the B-field of our NCOS theory has a non-zero flux
on the torus. We show that the NCOS, although being an open string theory
without any closed strings, enjoys the full  $SO(2,2;Z)\simeq
SL(2,Z)\times SL(2,Z)$ T-duality group.
This T-duality group combined with the S-duality $SL(2,Z)$ group yield
the usual U-duality group of closed strings on $T^2$; $SL(3,Z)\times
SL(2,Z)$ \cite{RS2}. 
We will argue that this U-duality group
can be realized from the OM theory \cite{OM} picture by a $T^3$
compactification.

\newsection{3+1 Noncommutative open strings, a systematic approach}

\def\lm{\left( \matrix{ }
\def\rm{ }\right) }

Let us consider a D3 brane in the presence of a constant $B$-field
with components $B_{01}$ and $B_{23}$.
Such constant $B$-field is equivalent to  constant electric
and magnetic vector fields on the brane  pointing in the direction $x^1$.
In our conventions $\mu, \nu =0,1,2,3$, label the directions 
parallel to the brane, and $x^a $, $a,b=4,...,9$, the transverse
directions.
It is convenient to choose coordinates so that the closed string metric
is of the form
\ba
g_{\mu\nu}&=&{\rm diag}( -\zeta,\zeta , \eta, \eta)\ , \cr
g_{ab}&=&K\delta_{ab}\ ,
\ea
where $\zeta, \ \eta $ and $K$  are constant parameters which will be
fixed later.
The boundary conditions for the open string coordinates are given by
\ba
& \partial_{\sigma}X^0+E \partial_{\tau}X^1 \ \bigg|_{\s=0,\pi }=0 \ ,\ \
\ \ 
\partial_{\sigma}X^1+E \partial_{\tau}X^0\ \bigg|_{\s=0,\pi }=0 \ , \\
& \partial_{\sigma}X^2+B \partial_{\tau}X^3\ \bigg|_{\s=0,\pi }=0 \ ,\ \ \
\ 
\partial_{\sigma}X^3-B \partial_{\tau}X^2\ \bigg|_{\s=0,\pi }=0 \ ,
\ea
$$
E\equiv  B^0_{\ 1}=- \zeta^{-1} B_{01}\ ,\ \ \ \ \
B\equiv B^2_{\ 3}=\eta ^{-1} B_{23}\ .
$$
The transverse open string coordinates are free string coordinates obeying
Dirichlet boundary conditions:
$$
X^a\ \bigg|_{\s=0,\pi }=x^a\ ,\ \ \ \ a=4,...,9\ ,
$$
where $x^a$ denote the position in transverse space of $N$ D3 branes.
The solutions to the above boundary conditions are
$$
X^0 =x^0+2\a ' (p^0\t - E p^1 \s ) +\sqrt{2\a '} \sum_{n\neq 0} {1\over
n}e^{-in\t }\big[ i
a^0_n
\cos(n\s
)- E a^1_n \sin(n \s) \big]\ , 
$$
$$
X^1 =x^1+2\a ' (p^1\t - E p^0\s ) +\sqrt{2\a '} \sum_{n\neq 0} {1\over
n}e^{-in\t }\big[ i
a^1_n
\cos(n\s
)- E a^0_n \sin(n \s)\big] \ , 
$$
\be
X^2 =x^2+2\a ' (p^2\t -B p^3\s )+\sqrt{2\a '} \sum_{n\neq 0} {1\over
n}e^{-in\t }\big[ ia^2_n
\cos(n\s
)- B a^3_n \sin(n \s)\big] \ , 
\label{OS}
\ee
$$
X^3 =x^3+2\a ' (p^3\t + B p^2\s )+ \sqrt{2\a '} \sum_{n\neq 0} {1\over
n}e^{-in\t }\big[ ia^3_n
\cos(n\s
)+ B a^2_n \sin(n \s)\big] \ , 
$$
$$
X^a=x^a+ \sqrt{2\a '}\sum_{n\neq 0} {\alpha _n^a \over n} e^{-i n\tau }
\sin (n\s )\ .
$$

   
\def\th{\theta }
\def\td{\tilde \theta }

Now we fix the parameters $\zeta$ and $\eta$ so that the {\it open string
parameters} $G_{\mu\nu},\
\theta_{\mu\nu}$   \cite{SW},
\be\label{open}
G_{\mu\nu}= g_{\mu\nu} - (B g^{-1} B)_{\mu\nu}\ , \ \ \ \ \
\theta^ {\mu\nu} = 2\pi \a' \left( {1\over g+ B }\right)^{\mu\nu}_A\ ,
\ee
get a simple form, convenient for describing the \NCOS{s}; i.e. we set 
\be\label{zeta}
\zeta =(1-E^2)^{-1}\ , \ \ \ \  \eta =(1+B^2)^{-1}\ ,
\ee
so as to have
\be
G_{\mu\nu}=\eta_{\mu\nu} \ , \ \ \  \ \th^{01}\equiv\theta=2\pi\alpha' E\
,\ \ \ \ \th^{23}\equiv\td=2\pi\alpha' B\ .
\ee
Then the open string coupling is given by
\be\label{Ocoup}
G_s=g_s \left({\det G_{\mu\nu}\over \det (g_{\mu\nu}+B_{\mu\nu })}
\right)^{1\over 2} = 
g_s \sqrt{(1-E^2)(1+B^2) }\ ,
\ee
The other parameter in the metric, $K$, can be fixed independently. 
A suitable way for fixing it is using the D3-brane RR charge
normalization. However, the way used in \cite{GMMS} is different.
Demanding this RR charge to remain the same before and
after the scaling of the closed string metric implies that
\be
K=\bigg({{\rm det}(g_{ab})\over {\rm det}(g_{\mu\nu})}\bigg)^{1/2}\ 
=[(1-E^2)(1+B^2)]^{-1/5}\ .
\ee

The canonical commutation relations for the string coordinates then imply the
following commutation relations for the mode operators:
$$
[x^\mu,x^\nu ]= i\theta ^{\mu\nu}\ ,\ \ \ [x^\mu,p^\nu ]=i G^{\mu\nu}\ ,
\ \ \ \ 
[a_n^\mu ,a_m^\nu ]=n \delta_{n+m} G^{\mu\nu}\ , \ \ \
[\a_n^a ,\a_m^b ]=n \delta_{n+m}K^{-1}\delta^{ab}\ .
$$
One can also check that {\it the end points} of the string do not commute,
i.e.
$$
[X^\mu (\tau, 0),X^\nu (\tau, 0)]=i \theta^{\mu\nu}\ ,\ \ \ \ \ 
[X^\mu (\tau, \pi),X^\nu (\tau, \pi)]=-i \theta^{\mu\nu}\ .
$$
The open string mass spectrum then becomes manifestly the same as
the free string mass spectrum:
$$
\a' M^2=\a' (p_0^2-p_1^2-p_2^2-p_3^2)= N-1\ .
$$
The addition of fermions is as in the usual free
open superstring theory (with the appropriate change in the
normal ordering constant).
The closed string mass spectrum is then given by
$$
\a' (1-E^2)  (p_0^2 -p_1^2 )- \a' (1+B^2) (p_2^2+p_3^2)= 2N+2\bar N-4\ .
$$
We see that as $E\to 1 $ the energy of the closed string states
goes to infinity, while the energy of open string excitations
remains finite.
In the limit $E\to 1 $ with fixed $B$, $\a '$ and $G_s$,
the closed string states are thus decoupled from the theory.
Note that this limit requires $g_s\to \infty $ (see eq.~(\ref{Ocoup})~)
and hence the closed string decoupling should be handled with more care;
moreover one should still discuss why the massless closed
string modes are also decoupled. This can be done noting that their
effective coupling, the ten dimensional Newton constant, $$
G_N={\a'}^4\ g_s^2\ K^{-5}(1-E^2)(1+B^2)={\a'}^4\ G_s^2 (1-E^2)(1+B^2)\ , 
$$  
in the critical $E$ limit goes to zero.

The resulting open string theory obtained in this
limit contains the parameters
$G_s $, 
\be
\theta= 2\pi \a'   \ ,\ \ \ \  \td =2\pi \a' B=\th B\ .
\label{uuuu}
\ee
Later we will show  that  one can also introduce 
another parameter $\chi$ associated with the RR scalar of type IIB theory.
So, altogether our NCOS theory is defined by four parameters,
$\alpha',\tilde\theta, G_s$ and $\chi $. 
The parameter $\alpha'= \theta /2\pi $ is the string scale and also characterizes the
noncommutativity scale in the $x^0$-$x^1$ directions, $\tilde \theta $
represents the noncommutativity scale in the
$x^2$-$x^3$ directions,  $G_s$ is the open string coupling, while $\chi $ is not relevant  in
the perturbative expansion.

General disc amplitudes for this ''noncommutative'' open string theory
will have the form
\be
\langle V(p^{1})...V(p^{N}) \rangle _{\theta,\td }
=\exp\big[ -{i\over 2} \sum_{n>m}
p^n\wedge p^m \epsilon(\tau_n-\tau_m)\big] \ 
\langle V(p^{1})...V(p^{N}) \rangle _{\rm free\ string}\ ,
\ee
where
$$
p^n\wedge p^m= (p_0^{n} p_1^{m}-p_1^{n} p_0^{m})
\th + (p_2^{n} p_3^{m}-p_3^{n} p_2^{m}) \td \ . 
$$
In the case $B=0$, one recovers the open string theory of
\cite{SST,GMMS}, obtained from open strings in a purely electric
background.

It is worth noting that the \NCOS s also carry $U(N)$ Chan-Paton  factors,
and therefore besides the parameters $(\a',\ G_s,\ \tilde\theta,\ \chi)$
in order to specify a NCOS completely we should also determine $N$.
We also note that in the $B\to \infty,\ \alpha'\to 0$, with
\be
G_s={\rm fixed}\ ,\ \ \ \ \alpha'B={\rm fixed}\ ,
\ee
limit of our NCOS theory we will recover the usual NCSYM theory, because
$\a'\to 0$,  massive open string excitations also decouple in this limit,
and one is left with
the Super Yang-Mills field theory in  noncommutative $x^2$-$x^3$ space.
Thus the present family of NCOS theories  interpolates  between 
the purely electric NCOS theory of \cite{SST,GMMS} and the NCSYM.

In general, the theory contains two energy scales, given by
$\theta $ and $\tilde \theta $, or $\a' $ and $B\a' $.
At distances  $L$ much larger than $\sqrt{\theta }$, $\sqrt{\td }$,
the theory reduces to ordinary SYM theory. For $B\gg 1$, there is a
regime $\theta \ll L^2 <\td $ in which the theory is  SYM field
theory on the noncommutative space $x^2$-$x^3$; string effects
can be ignored, but noncommutativity effects in $x^2$-$x^3$ directions are
important. If $B$ is of order 1 or lower, then string effects appear
at the same time as noncommutative effects.

\newsection{SL(2,Z) S-duality transformations on NCOS theory}

Now we study the behavior of the open string theories of
the previous section under type IIB $SL(2,Z)$ symmetry. 
Let us start with the corresponding type IIB supergravity solution.
The Lorentzian supergravity solution representing a D3 brane in the
presence of $B_{01}$ and $B_{23}$ fields is given  in \cite{RS1}.
In the string frame, it is 
\be
ds_{str}^2 =  f^{-1/2} \bigg[ h'(- d{x_0}^2 +
 d{x_1}^2) + h (d{x_2}^2 +d{x_3}^2)\bigg] 
+f^{1/2} \bigg[ dr^2 + r^2 d\Omega_5^2 \bigg]\ ,
\label{susu}
 \ee
$$
f  = 1 + { {\alpha'} ^2 R^4 \over r^4 } ~, ~~~~~
{h}^{-1} = \sin^2\alpha f^{-1}+ {\cos^2\alpha } \ , ~~~~~ 
{h'}^{-1} = -\sinh^2\beta f^{-1}+ {\cosh^2\beta } \ ,
$$
$$
B_{01} = - \tanh\beta f^{-1} h'  \ ,\ \ \ \ \ \
B_{23} =  \tan\alpha f^{-1} h  \ ,
$$
$$  
e^{2\phi} = {g_s}^2 h h'\ ,\ \ \ \ \
\chi = {1 \over g_s } \sinh\beta\sin\alpha f^{-1}\  +\chi_0, 
$$
$$
A_{01} = ({1 \over g_s} \sin\alpha \cosh\beta +\chi_0 \tanh\beta)
h'f^{-1} \ , \ \ \
A_{23} = ({1 \over g_s} \sinh\beta \cos\alpha -\chi_0 \tan\alpha) hf^{-1}
,
$$
$$
F_{0123u}  = {1 \over g_s}\cos\alpha \cosh\beta \  h h' \partial_r f^{-1}
\ .
$$ 
In the $r\to\infty $ region, the metric  asymptotically approaches the
Minkowski metric, and the
asymptotic  values for the different  fields
are as follows
\be
(B^{\infty})^{0}_{\ 1} = \tanh\beta \equiv E \ ,\ \ \ \ \ \
(B^{\infty})^{2}_{\ 3} =  \tan\alpha \equiv B  \ ,
\nonumber
\ee
\be
(A^{\infty})^0_{\ 1} =-{1\over g_s}{B\over \sqrt{1+B^2}}{1\over \sqrt{1-E^2}}-\chi_0 E  \ , \ \ \ \
\  
(A^{\infty})^{2}_{\ 3} ={1\over g_s}{1\over \sqrt{1+B^2}}{E\over \sqrt{1-E^2}}- \chi_0
B\ ,
\label{asym}
\ee
\be
e^{2\phi_{\infty}}=g_s^2\ ,\ \ \ \ \chi_{\infty}={1\over g_s}{B\over \sqrt{1+B^2}}{E\over
\sqrt{1-E^2}} +\chi_0\ .
\nonumber
\ee
We see that in the $E\to 1$ limit, with 
$G_s, \theta, \tilde\theta$ and $\chi_0$ fixed (see eqs.~(\ref{Ocoup}), (\ref{uuuu})~),
 $A_{01}^\infty$
and
$A_{23}^\infty $ also remain
finite. 

Under the $SL(2,Z)$ symmetry of the type IIB superstring
the coupling
$$
\lambda=\Theta + {i\over g_s}\ ,\ \ \ \ \ \Theta\equiv \chi_\infty
\ ,
$$
transforms as 
\be\label{sltau}
\lambda\rightarrow \lambda'={a\lambda+b\over c\lambda+d}\ ,\ \ \ \ ad-bc=1\ ,
\ee
where $a,b,c,d$ form an $SL(2,Z)$ matrix, whereas  $B_{\mu\nu}$ and
$A_{\mu\nu}$ (NSNS and RR) fields form a doublet:
\be
\left(\matrix{ B_{\mu\nu}\cr 
A_{\mu\nu}}\right)\rightarrow
\left(\matrix{ B'_{\mu\nu}\cr
A'_{\mu\nu}}\right)=
\left(\matrix{d   & -c\cr
-b & a }\right)
\left(\matrix{ B_{\mu\nu}\cr
A_{\mu\nu}}\right)\ .
\label{ddw}
\ee
Therefore
\be
g_s\rightarrow g_s'= g_s|c\lambda+d|^2\ .
\label{ggs}
\ee
The Einstein metric
$g^{E}_{\mu\nu}=e^{-\phi/2} g^{\rm str}_{\mu\nu}$ remains invariant,
so  the new string metric at $r=\infty $ is 
$|c\lambda+d|\eta_{\mu\nu}$. 
Using eqs.~(\ref{asym}),~(\ref{ddw}), one finds that 
the transformed electric and magnetic fields are  
\be\bea{cc}\label{E'B'}
E\rightarrow E'={1\over |c\lambda+d|}\big[ (d+c\Theta )E  +c {B\over G_s}(1-E^2)\big]\ ,
\\
B\rightarrow B'={1\over |c\lambda+d|}\big[  (d+c\Theta )B- c {E\over G_s}(1+B^2)\big]\ .
\label{xxf}
\eea\ee
Let us now consider the  $E\to 1$ limit
for the $SL(2)$ rotated parameters. 
In this limit the transformation (\ref{xxf}) simplifies, since
$g_s\to \infty \ ,\ \lambda\to \Theta $.
One can distinguish two different cases:

\noindent {\it a)} Irrational $\Theta$:

In this case there are no integers $c$ and $d$ such that $c\chi+d=0$.
In the 
$E\to 1$ and $g_s\to \infty$ limit, $|c\lambda+d|$ reduces to
$|c\Theta +d|$ and $E'$ and $B'$ are:
\be
E'={1\over |c\Theta +d|} (d+c\Theta)E =\pm 1\ ,
\ \ \ \ 
B'=\pm B-{c(1+B^2) \over G_s |c\Theta +d| }={\rm
finite}\ .
\ee
In the $E\to 1$ limit, the electric and magnetic fields are obtained by the simple transformation
rules  
$$
{1-{E}^2 \over |c\lambda +d |^2}
\to {1-{E'}^2 \over |c\Lambda+d|^2}\ , \ \ \ \ 
{1+{B}^2 \over |c\lambda +d|^2 }
\to {1+{B'}^2 \over |c\Lambda+d|^2}\ ,
$$ 
where
\be
\Lambda =\chi_0+ {i\over G_s}\ , \ \ \ \ \ \chi_0=\Theta -{B\over G_s}\ .
\ee
The fact that  ${E'}^2\to 1$, with $B'$ finite (and also 
$g'_s\sqrt{1-{E'}^2}$ remains finite) shows that 
an $SL(2,Z)$ transformation leads to another NCOS with 
transformed parameters
$$
(\theta,\td ,G_s,\Theta )\ \longrightarrow \ (\theta ',\td  ',G_s',\Theta ')\ ,
$$ 
where
\be
\Theta '= {a\Theta+b\over c\Theta+d}\ , \ \ \ \ \ 
G_s'=G_s |c \Lambda +d|^2\ .\ \ \
\ee
To find ${\theta'}^{\mu\nu}$ and $G'_{\mu\nu}$  we use
eq.(\ref{open}).
One can choose coordinates (by scaling by suitable factors $\zeta$ and $\eta$ given by
(\ref{zeta})
as in the previous section) so that the open
string metric before the $SL(2)$ transformation is
$G_{\mu\nu}=\eta_{\mu\nu} $. 
We find
\be
G'_{\mu\nu}= {|c\Lambda+d|^{2}\over |c\lambda+d|}\ \eta_{\mu\nu}\ ,
\ee
\be
\theta '=2\pi\a ' {E'(1-E^2)\over
|c\lambda+d| (1-{E'}^2)}
= 2\pi\a ' {|c\lambda+d|\over |c\Lambda+d|^2}\ ,
\ee
\be
\tilde \theta '=
2\pi\a ' {B'(1+B^2)\over |c\lambda+d| (1+{B'}^2)}=
2\pi\a 'B' {|c\lambda+d|\over |c\Lambda+d|^2}\ .
\ee
Rescaling the coordinates so that the new open string metric $G'_{\mu\nu}$
is equal to
$\eta_{\mu\nu}$, we conclude that the new NCOS theory has parameters,  
\be
\theta '=2\pi\a '\ , \ \ \ \ \tilde \theta '=2\pi\a '\ B'\ .  
\ee

\noindent {\it b)} Rational $\Theta$:

In this case there exists an $SL(2,Z)$ transformation under which
$c\Theta +d=0$.
The string coupling transforms as follows:
\be
g_s\rightarrow g'_s=g_s|c\lambda+d|^2={c^2\over g_s}\ ,
\ee
i.e. this transformation relates strong and weak coupling regimes.
{} From (\ref{E'B'}) one can find transformed $E$ and $B$ in the $E\to 1$
limit:
\be
E' = \pm {B\sqrt{1-E^2}\over \sqrt{1+B^2}}
\to 0\ , \ \ \ \ \
B'=\pm {\sqrt{1+B^2}\over \sqrt{1-E^2}}\rightarrow \pm \infty\ .
\ee
Therefore 
\be
\theta'=2\pi\alpha' {E'(1-E^2)\over |c\lambda+d| (1-{E'}^2)}=0\ , \ \ \ \
\tilde \theta '=
2\pi\a ' {B'(1+B^2)\over |c\lambda+d| (1+{B'}^2)}=2\pi\alpha' {G_s\over c}={\rm finite},
\ee
and the open string coupling is
\be
G'_s=g'_s\sqrt{1-{E'}^2}\sqrt{1+{B'}^2}=g'_s B'={c^2(1+B^2)\over G_s}={\rm finite}.
\ee

In conclusion, for rational $\Theta$ there is an $SL(2,Z)$ transformation
which maps
the  NCOS theory to  a NCSYM field theory, 
with   $\Theta={a\over c}$. 
In particular, we see that whenever $\Theta=0 $
the NCOS theory is
S-dual to NCSYM theory,
even if both $E$ and $B$ are non-vanishing. 
The reason is that in this case $A_{01}$ given in
(\ref{asym})
also vanishes, and therefore the S-dual theory will have a vanishing $B_{01}$.
This generalizes the result of \cite{GMMS} that the theory with $E=0,\ B\to\infty $ is S-dual to
the theory with $E=1,\ B= 0$.
For irrational $\Theta$, 
under $SL(2,Z)$ transformations 
the NCOS theory is always transformed to a  NCOS theory with
new parameters given
by the above transformation rules.

\newsection{NCOS Theory on a Torus and T-duality} 

Noncommutative super Yang-Mills theory on a general $T^2$ torus,
unlike its commutative counter-part, enjoys the full $SO(2,2;Z)$
T-duality group of the underlying string theory \cite{SW}.
In the gauge theory language this is due to the ``Morita equivalence",
which is an equivalence for the gauge bundles on the noncommutative
torus, with a proper mapping between the corresponding
gauge groups and couplings, background magnetic fluxes, and  volumes of
the two tori.

NCOS theory (with rational $\chi $) is equivalent to NCSYM theory by
S-duality. On the other hand T-duality, although being a perturbative
symmetry, should also hold in  the strong coupling limit. Hence 
the T-duality group of NCOS theory with two spatial dimensions
compactified on a 2-torus, must be same as the T-duality group of NCSYM
theory, $SO(2,2;Z)$. 

Let us now consider T-duality transformations on NCOS theories we
introduced previously. To study  T-duality of NCOS theories
is convenient to start with the appropriate type IIB configuration and
then take a limit leading to a NCOS theory.
Once the moduli parameters of the compactified theory are specified,
it is easy to obtain the  T-duality transformation properties.
For the case of our interest, (3+1) dimensional NCOS with parameters,
$(\a',\ G_s,\ \tilde\theta,\ \chi)$, we consider the
compactification of the $\theta$-plane $x_2$-$x_3$ on a noncommutative
torus two torus  $T^2_{\theta}$.
This theory can be realized as some particular limit of type IIB string
theory
in the presence of a $(D3,(F,D1))$ -brane
system compactified on the two torus.
We denote the complex and Kahler parameters of that torus by $\tau$ and
$\rho$, respectively. Then the brane bound state
is characterized by two integers $m,\ N$, whose ratio is proportional to
the RR charge density corresponding to
D-strings \cite{{AAS}}.
We choose coordinates so that the components of the
closed string metric parallel to the brane bound state are
$(-{1\over 1-E^2},\ {1\over 1-E^2},\ 1,\ 1)$.
Along the lines of  \cite{{Dip},{AAS}}, the spectrum of the open strings
attached to the brane bound state is
\be
\a' M^2={|r+q\tau|^2\over \tau_2}\ {\rho_2\over |m+N\rho|^2}+ {\rm
Oscil.}\ ,
\ee
where $\tau_2$ and $\rho_2$ are the imaginary parts of $\tau$ and $\rho$,
$r$ and $q$ are two integer parameters
representing the winding and momentum modes of open strings, respectively.
We see  that the
zero mode part of the spectrum is manifestly invariant
under the T-duality group $SO(2,2;Z)\sim SL(2,Z)_{\tau}\times\
SL(2,Z)_{\rho}$.
The other open string parameters are
\be
\theta^{01}=-\theta^{10}=2\pi\a' E\ \ \ \ \ \
\theta^{23}=-\theta^{32}=2\pi\a'{B\over 1+B^2}\ ,
\ee
\be
G_s=g_s\sqrt{(1-E^2)(1+B^2)}\ .
\ee
Now we take the $E\to 1$ limit while keeping $\a'$, $G_s$ and the
volume of the torus fixed. This  leads
to a NCOS theory on $T^2_{\theta}$ defined by parameters: 
$(\a',\ G_s,\ \tilde\theta,\ \chi; m ,\ N,\ R_1,\ R_2)$.
The $SL(2,Z)_{\tau}$ part consists of transformations under which
$\td,\ m,\ N $, $\chi$  and $ G_s$ are invariant; it only acts on
the torus metric (and $r$ and $q$ modes).
Other  transformations are generated by $SL(2,Z)_{\rho}$ matrices
which  act on the torus volume $V,\ \theta,\ G_s$ and $(m,\ N)$ as
\cite{SW}
\be 
V'=V\ (a+b\vartheta)^2\ \ \ , G'_s=G_s\ (a+b\vartheta)\ , \ \ \ \
\vartheta'={c+d\vartheta\over a+b\vartheta}\ ,
\ee
\be
\left(\matrix {m'\cr N'}\right)=\left(\matrix{a& b\cr c&  
d}\right)\left(\matrix{m \cr N}\right)\ ,\ \ \ \ ad-bc=1\ , 
\ee
where $\vartheta={\tilde\theta\over V}$. 
Thus, under this transformation, a NCOS theory is mapped into another
NCOS theory with the same $\a'$ and $\chi$
parameters, while all other moduli are transformed as above.
For the special case of rational $\vartheta$ there is a T-duality
under which $\vartheta$ vanishes and the resulting theory is NCOS theory
with $\tilde\theta=0$.

Type IIB string theory has in addition the strong-weak  $SL(2,Z)$ duality
symmetry. This S-duality combined with the above mentioned T-duality group
in the usual manner yields the U-duality group $SL(3,Z)\times SL(2,Z)$.

Since general 3+1 NCOS can be obtained from the  OM theory \cite{OM} on a
torus \cite{RS2}, the NCOS on compactified on a $T^2$ can also be
obtained as a limit of OM theory compactified on a 4-torus $T^4$.
And therefore it may seem that the U-duality symmetry group of NCOS
theory could be larger than the $SL(3,Z)\times SL(2,Z)$ group.
However, this is not the case,
because in order to obtain  NCOS theory in 3+1 dimensions (with no
additional Kaluza-Klein states coming  from $d=6$) one must take a zero
radius limit of $R_4, R_5$ which  breaks the
additional  symmetry. Indeed, after compactifying M-theory on  $T^2\times
T^2$ one has type IIB string theory in $d=7$, i.e. type IIB on
$R^7\times T^3$, which has a larger U-duality group, $SL(5,Z)$.
To have type IIB string theory in $d=8$, one needs to take an infinite
radius limit of one of the $S^1$ in $T^3$
(which, in M-theory variables, corresponds to the zero area limit of one
of the 2-torus). This reduces the U-duality group to $SL(3,Z)\times
SL(2,Z)$.

In the end I would like to mention two interesting open problems:

{\it i)} The NCOS strings, being a string theory, show the stringy
thermodynamics, in particular the Hagedorn transition \cite{Hage}. On the
other hand, the theories at strong coupling are related to  \nc gauge
theories. It would be quite interesting to see whether this phase
transition can also be realized in the field theory limit. There are some
motivations that it should. In particular we note that the monopole
solutions of \nc gauge theories, unlike the \com case, are not point-like
objects and they are stringy. So it seems that in the regime that the
dynamics is governed by monopoles, the strong coupling, one should be able
to see the stringy behaviour including Hagedorn transition.

{\it ii)} The \nc open strings we discussed here carry NC$U(N)$ Cahn-Paton
factors and hence they are oriented open string theories. It would be
very interesting to look for the formulation   of {\it unoriented} \NCOS
s. Although it may seem problematic at first sight, since NC $SO(N)$ and 
$Sp(N)$ gauge theories have been formulated \cite{NCSO}, the    
unoriented \NCOS s are those appear as the strong coupling limit of
these gauge theories \cite{UNCOS}.


{\bf Acknowledgements}

The warm hospitality of the organizers of conference {\it BRANE NEW WORLD
AND NONCOMMUTATIVE GEOMETRY} is acknowledged.
I would like to thank A. Schwarz for discussions.


\end{document}